# Universal scaling of the critical temperature and the strange-metal scattering rate in unconventional superconductors

Jie Yuan[1,2,7], Qihong Chen[1,3,7], Kun Jiang[1], Zhongpei Feng[1,3], Zefeng Lin[1], Heshan Yu[1], Ge He[1], Jinsong Zhang[1], Xingyu Jiang[1], Xu Zhang[1], Yujun Shi[1], Yanmin Zhang[1], Zhi Gang Cheng[1,3], Nobumichi Tamura[4], Yifeng Yang[1], Tao Xiang[1], Jiangping Hu[1,3*], Ichiro Takeuchi[5,6*], Kui Jin[1,3*], and Zhongxian Zhao[1,3]

[1]Beijing National Laboratory for Condensed Matter Physics, Institute of Physics, Chinese Academy of Sciences, Beijing 100190, China.
[2]Key Laboratory of Vacuum Physics, School of Physical Sciences, University of Chinese Academy of Sciences, Beijing 100049, China
[3]Songshan Lake Materials Laboratory, Dongguan, Guangdong 523808, China.
[4]Advanced Light Source, Lawrence Berkeley National Laboratory, Berkeley, CA 94720, USA
[5]Department of Materials Science and Engineering, University of Maryland, College Park, MD 20742, USA.
[6]Quantum Materials Center, University of Maryland, College Park, MD 20742, USA.
[7]These authors contributed equally.

Emails: jphu@iphy.ac.cn; takeuchi@umd.edu; kuijin@iphy.ac.cn

**Abstract: Dramatic evolution of properties with minute change in the doping level is a hallmark of the complex chemistry which governs cuprate superconductivity as manifested in the celebrated superconducting domes as well as quantum criticality taking place at precise compositions[1–4]. The strange metal state, where the resistivity varies linearly with temperature, has emerged as a central feature in the normal state of cuprate superconductors[5–9]. The ubiquity of this behavior signals an intimate link between the scattering mechanism and superconductivity[10–12]. However, a clear quantitative picture of the correlation has been lacking. Here, we report observation of quantitative scaling laws between the superconducting transition temperature $T_c$ and the scattering rate associated with the strange metal state in electron-doped cuprate $La_{2-x}Ce_xCuO_4$ (LCCO) as a precise function of the doping level. High-resolution characterization of epitaxial composition-spread films, which encompass the entire overdoped range of LCCO has allowed us to systematically map its structural and transport properties with unprecedented accuracy and increment**

**of $\Delta x$ = 0.0015. We have uncovered the relations $T_c \sim (x_c-x)^{0.5} \sim (A_1^\square)^{0.5}$, where $x_c$ is the critical doping where superconductivity disappears on the overdoped side and $A_1^\square$ is the scattering rate of perfect $T$-linear resistivity per $CuO_2$ plane. We argue that the striking similarity of the $T_c$ vs $A_1^\square$ relation among cuprates, iron-based and organic superconductors is an indication of a common mechanism of the strange metal behavior and unconventional superconductivity in these systems.**

**Main text:**

The strange metallic behavior in the normal-state resistivity of cuprate superconductors was first observed shortly after their discovery. The unusual behavior where the resistivity varies as a linear function of temperature (linear-in-$T$ resistivity) has now been reported in a number of superconducting copper oxides up to several hundred kelvin[5,6]. In a narrow composition region around optimal doping, the linear-in-$T$ behavior extends to low temperatures (close to $T_c$)[6], indicating a critical behavior at the quantum critical point (QCP). In a hole-doped copper oxide $La_{2-x}Sr_xCuO_4$ (LSCO), the linear-in-$T$ resistivity was found to dominate the normal-state transport down to 1.5 K in pulsed high magnetic fields, within an extended range near the optimal doping[7]. For electron-doped copper oxides, a perfect linear-in-$T$ resistivity was found to persist down to 40 mK in $Pr_{2-x}Ce_xCuO_4$[8] and to 20 mK in $La_{2-x}Ce_xCuO_4$ (LCCO)[9]. In particular, this strange metal behavior in LCCO was found to start at the doping level associated with the Fermi surface reconstruction ($x \approx 0.14$) to the endpoint of the superconducting dome. Moreover, the scattering rate of linear-in-$T$ resistivity (i.e., the coefficient $A_1$ from $\rho = \rho_0 + A_1T$) shows a positive correlation with $T_c$, suggesting the anomalous normal state and the superconductivity have the same origin[9,10].

There has been a concerted effort in the community to quantify the relationship between $A_1$ and $T_c$ as a direct function of the chemical doping concentration[10,11]. However, due to the lack of sufficient data points with enough density to map across the doping phase diagrams, an explicit expression had been elusive. To this end, we have employed high-precision thin-film composition spreads encompassing the entire concentration range of the electron-doped superconducting dome with incremental accuracy in doping concentration $\Delta x$ of 0.0015. The systematic measurements have uncovered a remarkable scaling law linking the superconducting transition temperature ($T_c$), doping level ($x$) and the $T$-linear coefficient ($A_1^\square$), namely $T_c \sim (x_c-x)^{0.5} \sim (A_1^\square)^{0.5}$, observed here for the first time. Our finding points to a striking universal relation between the normalized $T$-linear coefficient and $T_c$ among cuprates, pnictides, and a class of organic superconductors, strongly suggestive of a common underlying physics at work in these unconventional superconductors.

For the family of electron-doped copper oxide superconductor LCCO, the superconducting phase is the so-called $T'$ phase, which is only stable in the form of thin films. Upon doping, LCCO evolves from an antiferromagnetic (AF) state to a superconducting phase with a dome-shaped phase diagram[3]. At the endpoint of the superconducting region ($x_c \approx 0.175 \pm 0.005$), it enters a metallic (non-superconducting) Fermi liquid state[11], and in the vicinity of this QCP ($0.14 < x < 0.17$), a strange metal state appears in the low-temperature limit (down to 20 mK) upon suppressing the superconductivity with magnetic fields[9]. Very recently, the strange metal state has also been observed in the antiferromagnetic regime (e.g. $x = 0.12, 0.13$)[13].

This calls for comprehensive mapping of the subtle concentration-dependent properties across this narrow composition range. Because of the relatively complex synthesis process, it has been nontrivial to tune the composition of LCCO films with high precision. We upend this challenge by enlisting the composition spread synthesis combined with micron-scale systematic characterization. Employing combinatorial laser molecular beam epitaxy[14,15], we have fabricated continuous composition spread LCCO films having a linear gradient in the Ce content along a predesigned sample direction with demonstrated composition variation accuracy of $\Delta x = 0.0015$ across the superconducting dome.

The synthesis scheme is shown in Fig. 1a. Two targets with nominal compositions of $La_{1.90}Ce_{0.10}CuO_4$ and $La_{1.81}Ce_{0.19}CuO_4$ are used as two ends of the composition spreads, corresponding to the optimal doping ($x = 0.10$) with highest $T_c$ and the metallic Fermi liquid state ($x = 0.19$), respectively. A series of unitcell-thick gradient wedges are deposited with the two targets in an alternating manner using moving mechanical shutters at 700 °C on a SrTiO$_3$ (STO) (100) substrate. This results in a $c$-axis oriented epitaxial composition spread LCCO thin film with continuously varying composition between La$_{1.90}$Ce$_{0.10}$CuO$_4$ and La$_{1.81}$Ce$_{0.19}$CuO$_4$ and uniform total thickness across the 10 mm length of the substrate. This growth technique ensures that the synthesis conditions are identical for the entire doping range.

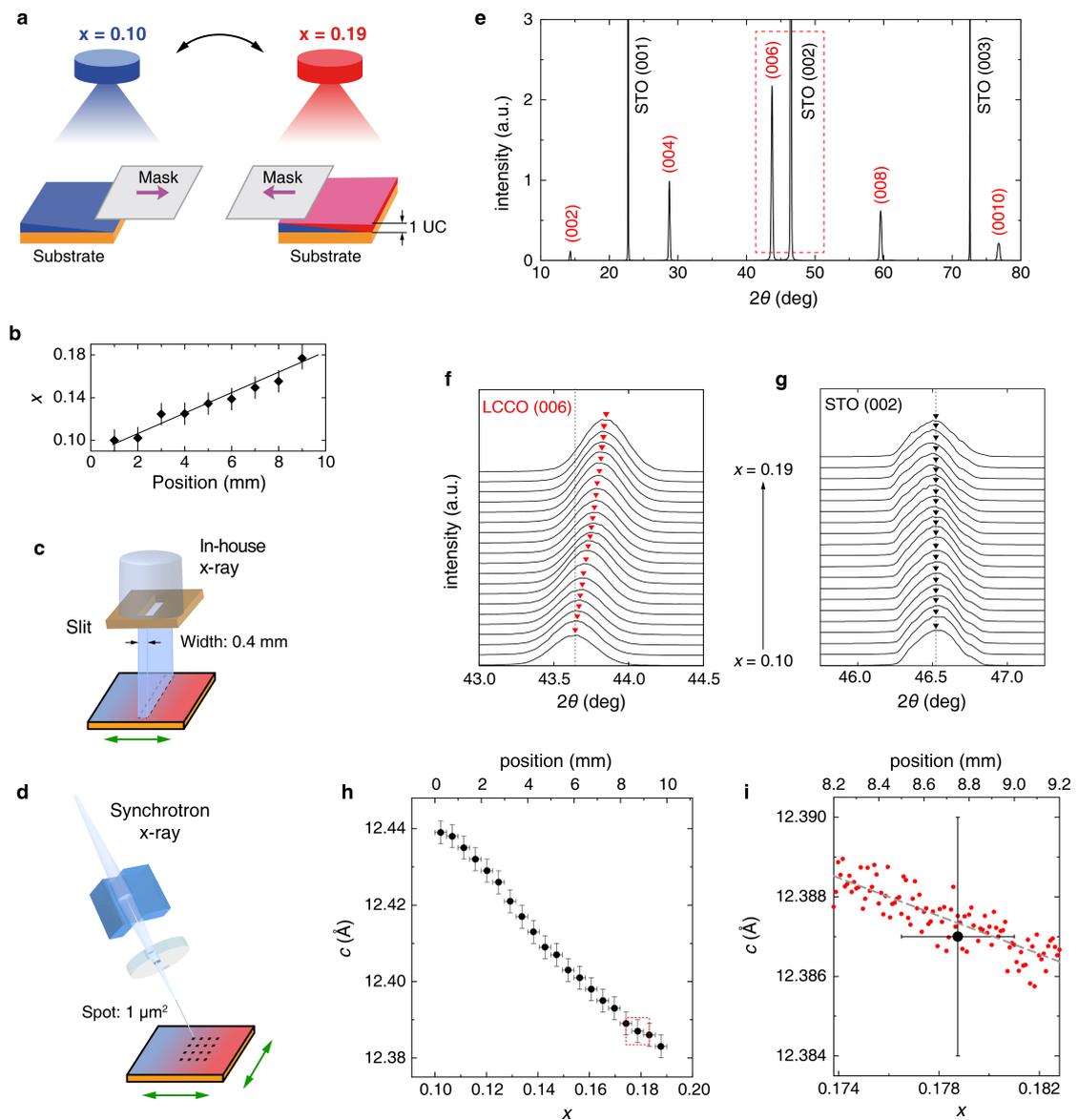

**Figure 1 | Combinatorial synthesis and multi-scale structural characterization of electron-doped cuprate superconductor La$_{2-x}$Ce$_x$CuO$_4$ (LCCO). a,** Schematic illustration of the composition-spread epitaxial growth of LCCO on a SrTiO$_3$ (STO)

single crystal substrate. Two targets with end compositions of the spread are ablated alternately, during which a shadow mask moves in such a way to create unit-cell-thick linear thickness gradients in opposite directions. A desired total thickness of the film (uniform across the spread) can be achieved by controlling the number of thickness gradient pairs to be deposited. UC: unit cell. **b,** The Ce doping level mapped across a spread with wavelength dispersive spectroscopy. **c,** The schematic illustration of the in-house x-ray diffraction (XRD) measurement configuration, with a beam width of 0.4 mm. **d**, The schematic of the synchrotron micro-diffraction, with a beam spot size of 1 μm$^2$. **e,** The $\theta/2\theta$ XRD obtained by in-house diffractometer with mm scale beam spot. **f-g,** The individual XRD patterns of the magnified $2\theta$ region (corresponding to the dashed box region in **e**), showing the (**f**) LCCO (006) and (**g**) STO (002) peaks. The spot size of the in-house x-ray is 0.4 mm, and the step for each line is 0.5 mm. The lines from different compositions are vertically shifted for clarity. The vertical dashed lines are guide to discern the peak shift. The inverted triangles mark the peak positions of each line. **h,** The doping dependence of the *c*-axis lattice constant across the spread measured by in-house XRD setup. The top axis shows the corresponding position dependence. **i,** The doping dependence of the *c*-axis lattice constant obtained by synchrotron diffraction (1 μm$^2$ spot) (red) from the region depicted by the red box in panel **h**. One lattice constant measurement (and error bars) obtained in this range in **h** is shown for comparison (black). The grey dashed line is the linear fitting for estimation of the uncertainty.

We first perform standard "low resolution" analysis of composition and the corresponding *c*-axis lattice constant variation across the spread using wavelength-dispersive x-ray spectroscopy (WDS) and an in-house diffractometer, respectively. As shown in Fig. 1b, the WDS-mapped Ce concentration in the LCCO film shows the expected dependence on position, spanning $0.10 \leq x \leq 0.19$. The uncertainty in determined concentration from WDS is typically 2%. Figure 1e shows the $\theta/2\theta$ x-ray diffraction (XRD) pattern from the entire spread integrated along the direction of compositional gradient. Figures 1f and 1g show the LCCO (006) and STO (002) peaks mapped along the spread, respectively. The LCCO (006) peak moves to higher angles as the doping concentration is increased, whereas the STO (002) peak does not change. Figure 1h shows the lattice constant mapping across the spread chip with error bars determined by limitations of the in-house diffractometer with a beam size of 0.4 mm operated under standard conditions. The smooth and well-behaved overall variation of the composition and the lattice constant over the entire length of the spread is thus confirmed despite relatively large measurement uncertainties associated with in-house characterization.

In order to harness the wealth of information which resides in the epitaxial spread at high spatial resolution, we enlist synchrotron microbeam diffraction (Fig. 1d) whose 1 micron beam spot size (together with micron-level accuracy of its scanning stage) allows the ultimate determination of the composition variation sensitivity and the smallest meaningful increment with which we can extract composition-dependent properties from the spread. Figure 1i shows the lattice constant mapping obtained from a part of the spread using the microbeam with 10 micron position increment across the spread (taken at Beamline 12.3.2 at Advanced Light Source). The variation of the lattice constant from point to point as well as its linear regression indicates that it can be determined with the uncertainty of 0.001 Å, which corresponds to the compositional variation $\Delta x$ of 0.0015. We note that such accuracy in composition control is not possible with traditional chemical synthesis methods[9]. Since positions on the spread can be readily specified with accuracy down to microns, these numbers ensure that we can obtain statistically significant mapping of composition-dependent properties with high incremental density within the 10 mm length of the spread encompassing $\Delta x \approx 0.09$.

The transport properties are obtained by patterning the spread film into micro-bridge arrays, as schematically shown in Fig. 2a. Initially, the entire LCCO spread film is patterned into 8 bridges, each with a width of 1 mm, for rapidly surveying the superconducting properties over the spread. Each bridge is then divided into 8 smaller bridges, each with a width of 100 μm. At the last step, some bridges with compositions near where the superconductivity disappears have been patterned into 20 μm wide bridges in order to study the delicate critical behavior.

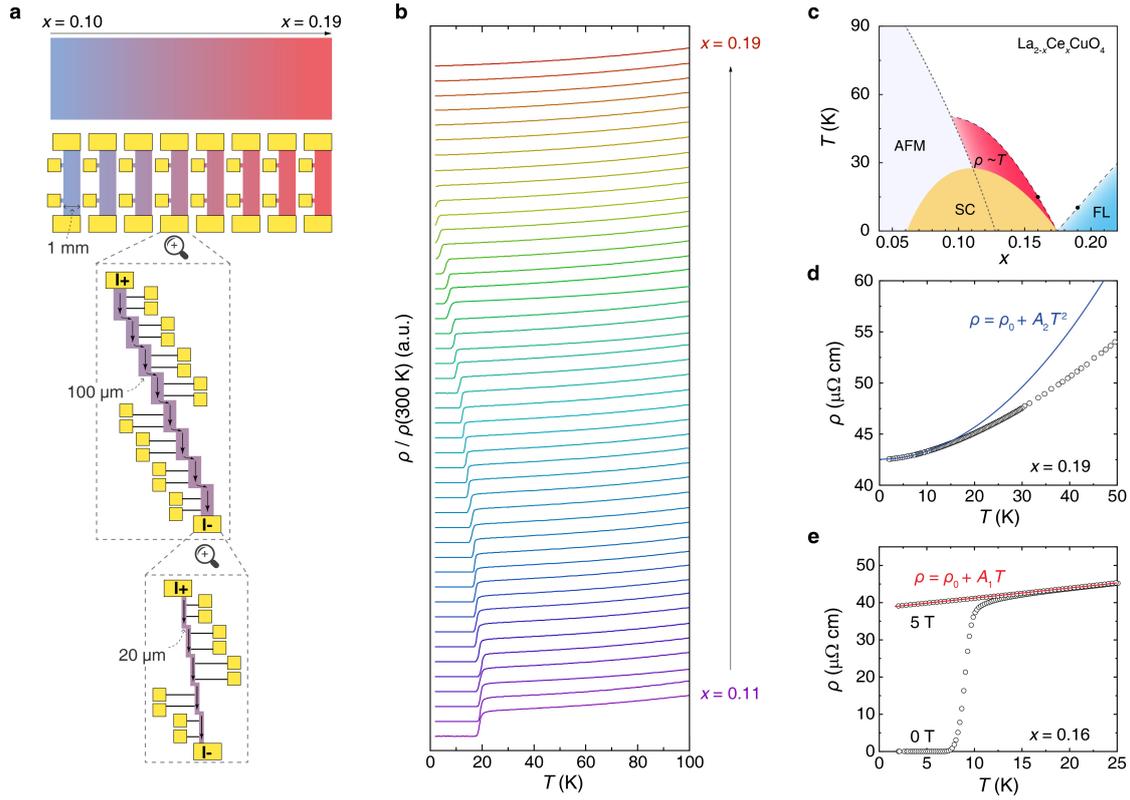

**Figure 2 | Micro-region characterizations of electrical transport properties. a**, Configurations of the patterned bridges for transport measurements across the spread. There are three levels of pattern arrays with successively decreasing microbridge widths: 1 mm, 100 μm and 20 μm from the widest to the narrowest. In this way, the spatial resolution of the transport measurement can reach the length scale comparable to the synchrotron microdiffraction mapping. **b,** Temperature dependence of the resistivity ($\rho$-$T$) for 100 micron-wide bridges patterned across the spread. The resistivity is normalized by its value at 300 K, and the curves are vertically shifted for comparison. From the bottom to top, the doping varies from $x = 0.10$ to $x = 0.19$. **c,** The phase diagram of LCCO. AFM, SC, and FL are short for antiferromagnetism, superconductivity, and Fermi liquid, respectively. Two representative dopings $x = 0.16$ and $x = 0.19$ are chosen for demonstration of the strange metal and Fermi liquid behaviors, respectively. **d,** Temperature dependence of the resistivity for $x = 0.19$. The solid line is the fitting using the Fermi liquid formula: $\rho = \rho_0 + A_2 T^2$. **e,** Temperature dependence of the resistivity for a bridge with $T_c \sim 10$ K ($x = 0.160$) at $B = 0$ T, and with a magnetic field of $B = 5$ T applied perpendicular to the film. The red straight line is a linear fitting $\rho = \rho_0 + A_1 T$.

The $R(T)$ data obtained from all 100-μm bridges across the entire spread are shown in Fig. 2b. Near the lower $x$ end (≈ 0.10), LCCO shows superconductivity with the highest $T_c$ ≈ 24 K (the bottom curve). Here, $T_c$ is defined as the temperature where the superconducting transition commences, as illustrated in Supplementary Fig. S2. Determining $T_c$ using different criteria does not influence any of our analysis (Supplementary Fig. S3). With increasing doping level, $T_c$ gradually decreases, and eventually bridges with higher Ce concentration only show the metallic behavior, where the resistivity decreases with decreasing temperature without any abrupt drop of resistance down to the lowest measured temperature of 2 K. The critical composition at which the superconductivity disappears corresponds to the doping level of $x_c$ = 0.177, consistent with previous results[9,16]. For $x > x_c$, the low-temperature dependence of resistivity obeys the Fermi liquid behavior, namely $\rho = \rho_0 + A_2 T^2$ (Fig. 2d), and the $T^2$ dependence persists to higher temperatures with increasing Ce content[9]. For $x < x_c$, the normal-state resistivity shows the linear-in-$T$ behavior at low temperatures which is ubiquitous in copper oxide superconductors. With the superconductivity suppressed by magnetic fields (applied perpendicular to the film), the linear-in-$T$ region extends down to the lowest measured temperature, where the experimental data can be fitted well by $\rho = \rho_0 + A_1 T$ (Fig. 2e).

The bottom panel of Fig. 3a shows the doping dependence of $T_c$ for two LCCO composition spread films. Compared with the limited data points obtained from samples made by the traditional synthesis method (blue squares, adapted from Ref [9]), a clear trend emerges in the dense data from the present combinatorial technique: the dashed line outlining the boundary of the superconducting phase obeys the square root relation $T_c \propto (x_c - x)^{0.5}$. The square root dependence has previously been used to fit the data of the hole-doped superconductor LSCO in the overdoped region[17]. Here, we are able to clearly discern it for the electron-doped LCCO in the overdoped superconducting regime for the first time.

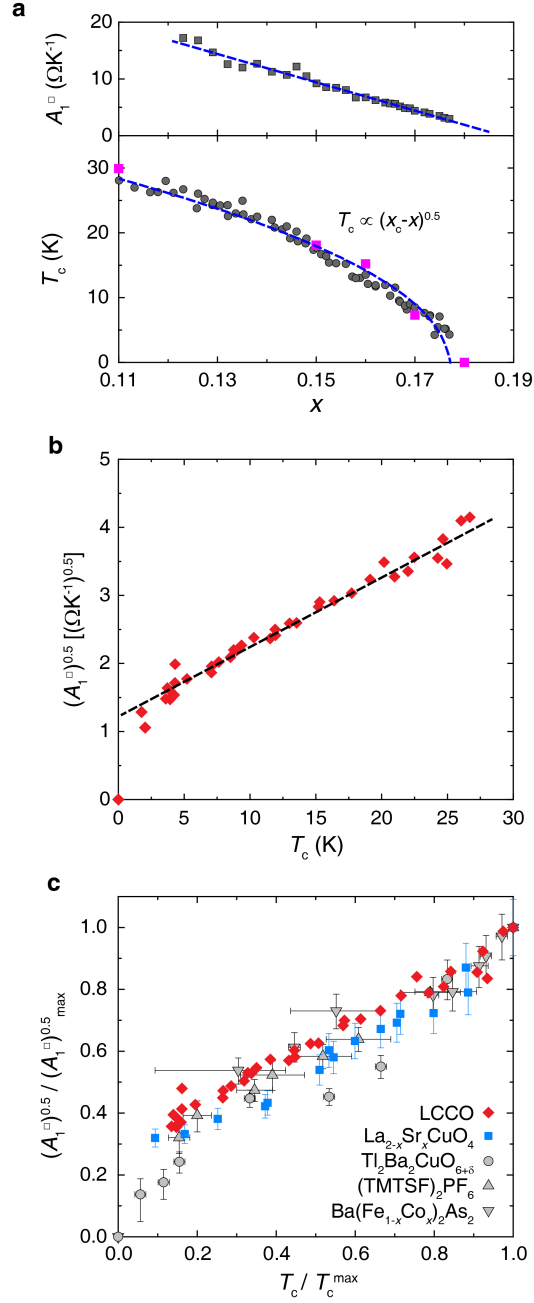

**Figure 3 | Quantitative scaling revealed from the systematic spread data and comparison of different unconventional superconductors. a**, Doping dependence of $T_c$ (bottom) and $A_1^{\square}$ (top). Across the full spread film doping range, $T_c$ exhibits a square root dependence on doping: $T_c \propto (x_c-x)^{0.5}$, while $A_1^{\square}$ shows a clear linear dependence on doping. The $T_c$ data are collected from two samples. The purple squares are extracted from Ref. [9]. **b,** $(A_1^{\square})^{0.5}$ as a function of $T_c$ extracted from the $R$-$T$ curves. The dashed straight line shows the linear fitting. **c,** The correlation between $(A_1^{\square})^{0.5}$ and $T_c$ for different superconducting systems from literature, with $(A_1^{\square})^{0.5}$ and $T_c$ normalized by their respective maximum values. LSCO, TBCO, TMTSFPF and BFCA are short for $La_{2-x}Sr_xCuO_4$, $Tl_2Ba_2CuO_{6+\delta}$, $(TMTSF)_2PF_6$ and

Ba(Fe$_{1-x}$Co$_x$)$_2$As$_2$, respectively. LSCO data are extracted from Ref. [20]; Data for (TMTSF)$_2$PF$_6$, Tl$_2$Ba$_2$CuO$_{6+\delta}$ and Ba(Fe$_{1-x}$Co$_x$)$_2$As$_2$ are from Ref. [10] and references therein.

Now we turn to the strange metal state in LCCO. We have extracted the $T$-linear coefficient $A_1$ from all 100-micron bridge curves shown here in the top panel of Fig. 3a as a function of the continuous doping level. As introduced by Legros et al.[18], we normalize $A_1$ by the distance between adjacent CuO$_2$ planes, i.e. $A_1^\square = A_1/d$, where $d$ is half of the $c$-axis lattice constant mapped accurately across the spread. An unmistakable linear dependence of $A_1^\square$ on doping ($x$) emerges as a result. We note that prior to this work, the relation between $A_1$ and the doping level had not been unambiguously quantified due to serious scattering of data points[10,19]. Without the combinatorial approach, it is difficult to obtain such accurate and systematic data. The newly unearthed relations in turn then immediately point to the square root dependence of $T_c$ on $A_1^\square$: Fig. 3b shows $T_c$ vs $(A_1^\square)^{0.5}$ with the dashed line given by the linear fit $(A_1^\square)^{0.5} = \alpha T_c + \beta$, with $\alpha = 0.10$ $(\Omega K^{-1})^{0.5}/K$ and $\beta = 1.22$ $(\Omega K^{-1})^{0.5}$. Hence, a linear relation between $(A_1^\square)^{0.5}$ and $T_c$ is established in LCCO.

This relation allows us to make quantitative comparisons with other unconventional superconductor systems. For the typical hole-doped copper oxide LSCO, the $A_1^\square(T_c)$ relation is extracted from a comprehensive study by Bozovic et al.[20]. As shown in Fig. 3c, the $A_1^\square$ shows a similar dependence on $T_c$. Beyond cuprates, the relation between $A_1^\square$ and $T_c$ has also been observed in the single-band organic superconductor (TMTSF)$_2$PF$_6$ as well as in the iron-based superconductor Ba(Fe$_{1-x}$Co$_x$)$_2$As$_2$. As seen in Fig. 3c, there is a single scaling relation which captures the common behavior among the disparate unconventional superconductors, suggestive of a universal underlying excitation governing their pairing mechanism.

This universal scaling relation places an explicit constraint on the general theory of linear-in-$T$ resistivity and unconventional superconductivity. Various theoretical descriptions[4,12,21–24] have been proposed to explain the linear-in-$T$ resistivity in copper oxides to date. One scenario attracting much attention involves the Planckian dissipation[18,24], where the scattering rate reaches the fundamental Planckian limit given by $\hbar/\tau = k_B T$. It can be used to explain the linear-in-$T$ resistivity in hole-doped copper oxides from high temperatures to the lowest measured temperature of 2 K. From $\Delta\rho = \rho - \rho_0 = m^*/ne^2\tau$ and the Planckian dissipation $\hbar/\tau = k_B T$, we obtain $\Delta\rho = \frac{m^* k_B}{ne^2 \hbar} T$. Previous studies have shown that in high-$T_c$ cuprates, the superfluid phase stiffness goes as $\rho_s \propto T_c$ and the superfluid density varies as $n_s \propto n$ (refs 20,25). These relations lead to $A_1^\square \propto m^*/n \sim m^*/n_s \sim (\rho_s)^{-1} \sim (T_c)^{-1}$, which is clearly in disagreement with the present observation of the positive correlation between $A_1^\square$ and $T_c$. This discrepancy needs to be resolved before the Planckian scenario can be applied to the overdoped side of copper oxides.

Another plausible origin of the linear-in-$T$ resistivity is the AF spin fluctuations associated with quantum criticality[3,23,26,27]. This picture is best substantiated in (TMTSF)$_2$PF$_6$, where superconductivity arises from a spin density wave (SDW) order and a perfect linear-in-$T$ resistivity is observed as $T$ approaches zero[10,28,29]. Its behaviors including the absence of a pseudogap and other anomalous electronic phases are believed to be intimately tied to short-range SDW fluctuations[3,10], the dominant source of transport scattering at low temperatures[30]. Very similar transport properties and evolution of ground states in the phase diagrams between LCCO and (TMTSF)$_2$PF$_6$, may be an indication that AF spin fluctuations are also at work in electron-doped cuprates[9,11]. Analogous behaviors have also been seen in the iron-based superconductor Ba(Fe$_{1-x}$Co$_x$)$_2$As$_2$[31]. Although the picture is more complex in hole-doped copper oxides involving a complex pseudogap and intertwined orders[32], the single scaling relation observed here might be the common signature of interplay among linear-in-$T$ resistivity, pairing correlations and spin fluctuations[10,30]. A microscopic description of how the pairing is mediated by spin fluctuations remains an open question, but given the universal behavior observed across different families of superconductors, renewed and focused theoretical investigations are perhaps in order.

In Fig. 3b, the linear fitting of $(A_1^\square)^{0.5}$ extrapolates to a finite value at $T_c = 0$. However, approaching the QCP at the end of the superconducting dome where $T_c = 0$, the linear-in-$T$ resistivity disappears, i.e. $A_1$ becomes zero[9]. This deviation is possibly due to quantum fluctuations or strong pairing fluctuations[33] near QCP, leading to the deviation of $(A_1^\square)^{0.5}$ from the linear dependence. Unfortunately, the uncertainty in $T_c$ determination and the reduced temperature range for linear resistivity close to QCP preclude us from obtaining a quantitative picture in this region. Further investigations are ongoing in search of further insight into the origin of superconductivity in the overdoped side of LCCO.

**Methods**

**Film growth.** We have fabricated $La_{2-x}Ce_xCuO_{4\pm\delta}$ ($x$ = 0.10 ~ 0.19) composition-spread thin films on $SrTiO_3$ (STO) substrates (10×10 mm$^2$ in size) with a programmable moving shadow mask. Two targets with compositions $La_{1.9}Ce_{0.1}CuO_{4\pm\delta}$ (i.e., $x$ = 0.10) and $La_{0.81}Ce_{0.19}CuO_{4\pm\delta}$ (i.e., $x$ = 0.19) are ablated alternately by ultraviolet KrF laser pulses ($\lambda$ = 248 nm). During the deposition, a moving mask with constant speed is used in order to generate opposing thickness wedges from the two targets. The deposition ratio was carefully controlled to ensure that the film deposited in one pair of wedge depositions never exceeded a single-cell layer in order to avoid formation of superlattices. After eighty periods, a 100 nm thick combinatorial $La_{2-x}Ce_xCuO_{4\pm\delta}$ (LCCO) film is fabricated followed by an *in situ* reduction process of several minutes in vacuum at about 700 °C.

**Structural characterization.** The crystallinity of the entire film is characterized by an in-house x-ray diffractometer with a slit, so that the beam spot size is 0.4 mm in width, and it is set to scan the spread film from one end to the other in the $2\theta$ range of 10° to 80°. The crystal structure of the spread film is also examined by synchrotron microdiffraction at Advanced Light Source (Beamline 12.3.2) with a beam spot of 1 μm$^2$. The sampling interval was 10 μm along the doping gradient direction (horizontal) for microdiffraction, and several points were also measured at each horizontal position with a vertical direction interval of 20 μm.

**Transport Measurements.** The composition-spread thin films were patterned into small chips, to take electrical resistivity from 300 K to low temperatures. There are three levels of device array patterns with successively smaller microbridge widths, and each set of measurements is followed by next level patterning and measurements. In this manner, the highest spatial resolution of the transport measurement is on the scale of 10 μm, comparable to the micron-resolution of synchrotron structural characterization.